\documentclass[a4paper,10pt]{article}

\usepackage[utf8]{inputenc}
\usepackage[english]{babel}
\usepackage{amsmath,amssymb}
\usepackage{enumitem}
\usepackage{todonotes}
\usepackage[natbibapa]{apacite} 
\usepackage{authblk}
\usepackage{multicol}
\usepackage{multirow}
\usepackage{placeins}

\DeclareMathOperator*{\med}{med}

\title{Machine Learning on EPEX Order Books: \\ Insights and Forecasts}
\author{S. Schnürch\thanks{Corresponding author: \url{simon.schnuerch@itwm.fraunhofer.de}} }
\author{A. Wagner}%\thanks{\url{andreas.wagner@itwm.fraunhofer.de}}}
\affil{\small Fraunhofer ITWM, Fraunhofer Institute for Industrial Mathematics ITWM, Fraunhofer-Platz 1, 67663 Kaiserslautern, Germany}

\date{\today}

\begin{document}
\maketitle              % typeset the header of the contribution
\begin{abstract}
\noindent This paper employs machine learning algorithms to forecast German electricity spot market prices.
The forecasts utilize in particular bid and ask order book data from the spot market but also fundamental market data like renewable infeed and expected demand.
Appropriate feature extraction for the order book data is developed.
Using cross-validation to optimise hyperparameters, neural networks and random forests are proposed and compared to statistical reference models.
The machine learning models outperform traditional approaches.

\vspace{1em}

\end{abstract}
\section{Introduction}
Forecasting electricity prices is an important task in an energy utility and needed not only for proprietary trading but also for the optimisation of power plant production schedules and other technical issues.
A promising approach in power price forecasting is based on a recalculation of the order book using forecasts on market fundamentals like demand or renewable infeed. 
However, this approach requires extensive statistical analysis of market data.
In this paper, we examine if and how this statistical work can be reduced using machine learning.
Our paper focuses on two research questions:
\begin{itemize}
\item How can order books from electricity markets be included in machine learning algorithms?
\item How can order-book-based spot price forecasts be improved using machine learning?
\end{itemize}

\begin{figure}[tb]
\includegraphics[width=.49\textwidth]{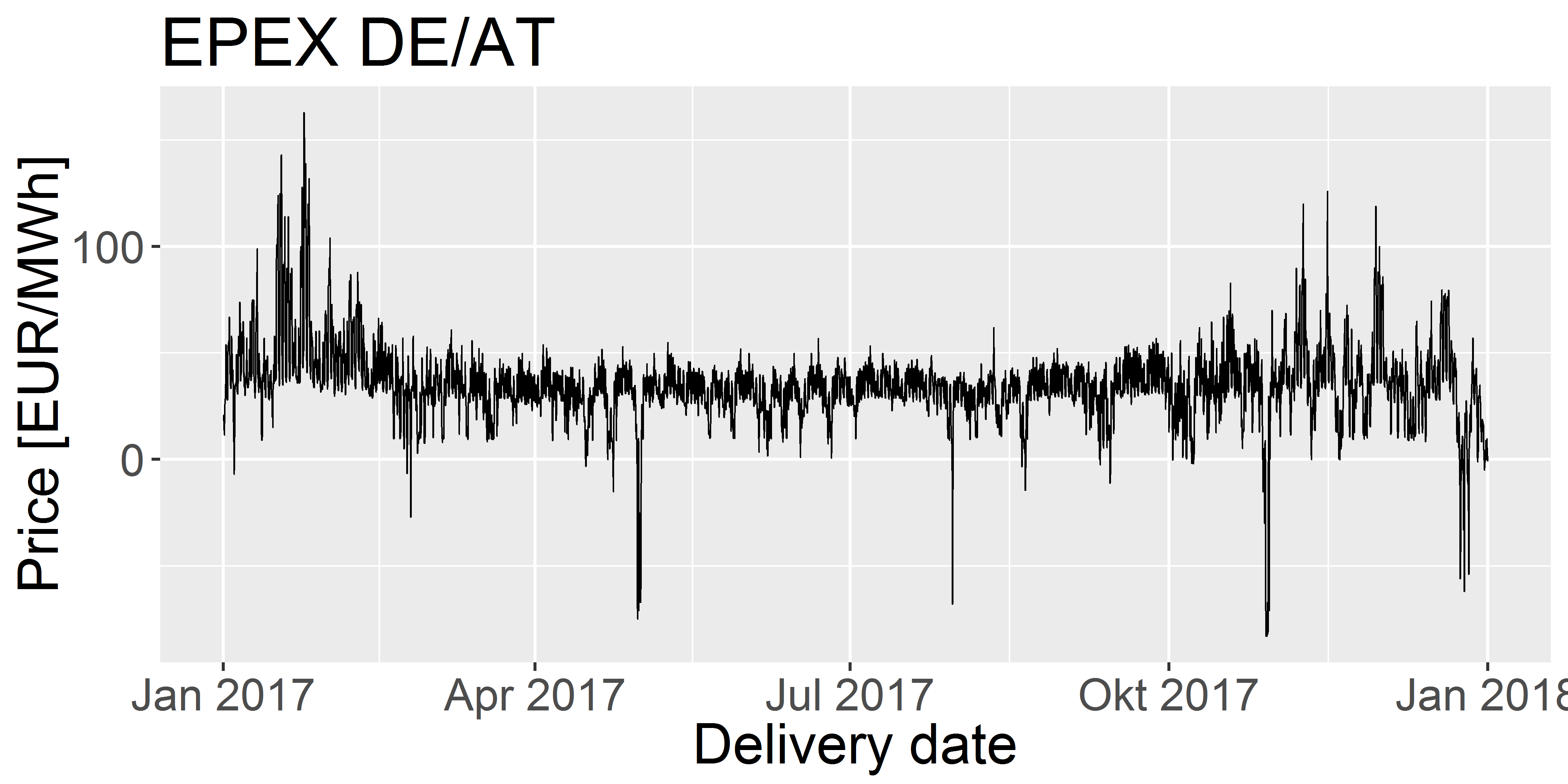}
\includegraphics[width=.49\textwidth]{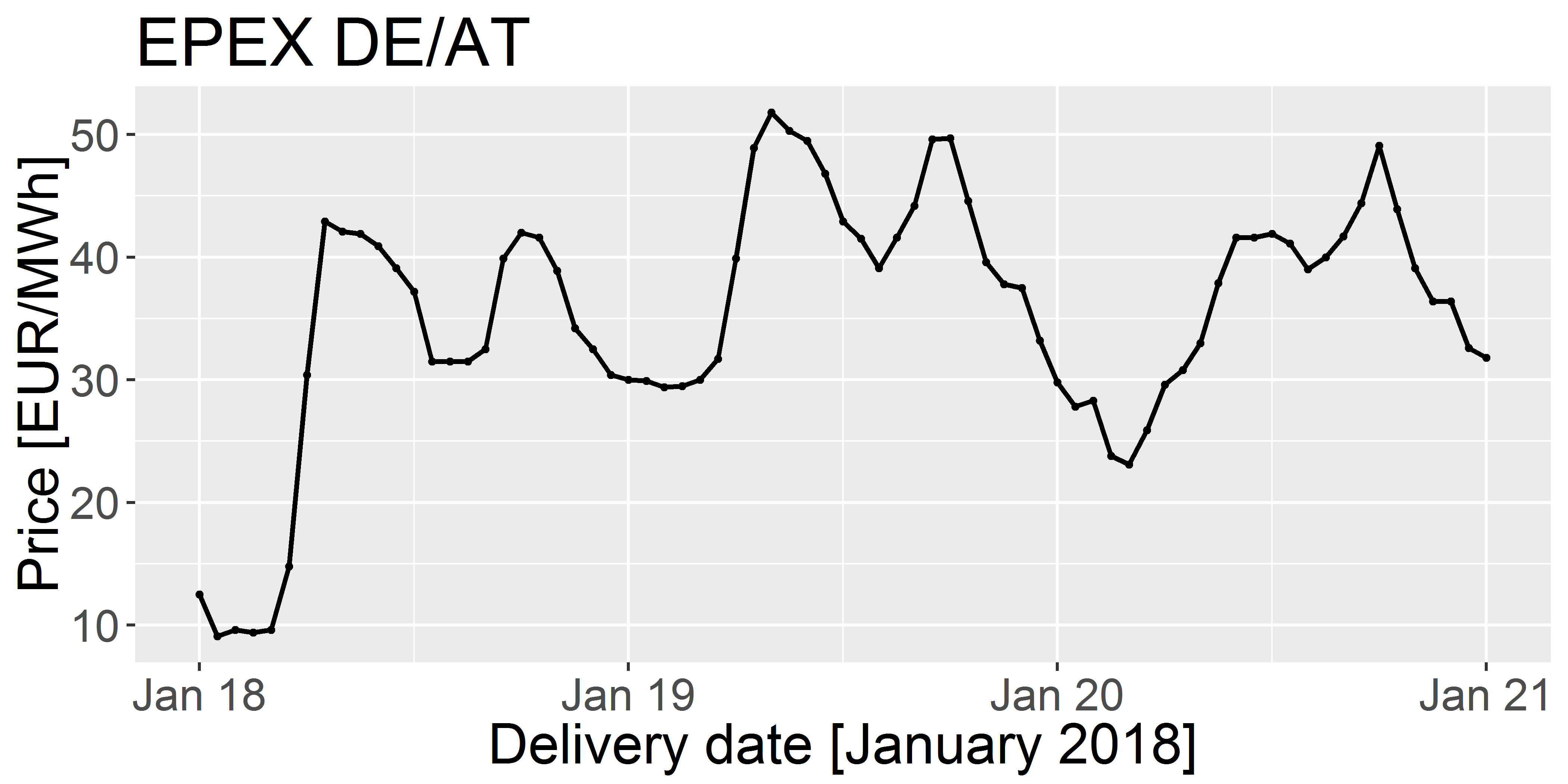}
\caption{Example electricity price time series on different time scales.}
\label{fig:prices}
\end{figure}

We consider the German/Austrian EPEX spot market for electricity.
There is a daily auction for electricity with delivery the next day.
All 24 hours of the day are traded as separate products.
Figure~\ref{fig:prices} shows auction results on different time scales.
The pronounced seasonality of prices is visible as well as their high volatility.\footnote{Another interesting property is that in contrast to price series of other commodities or stocks, electricity prices may become negative.}

\begin{figure}[tb]
\includegraphics[width=.49\textwidth]{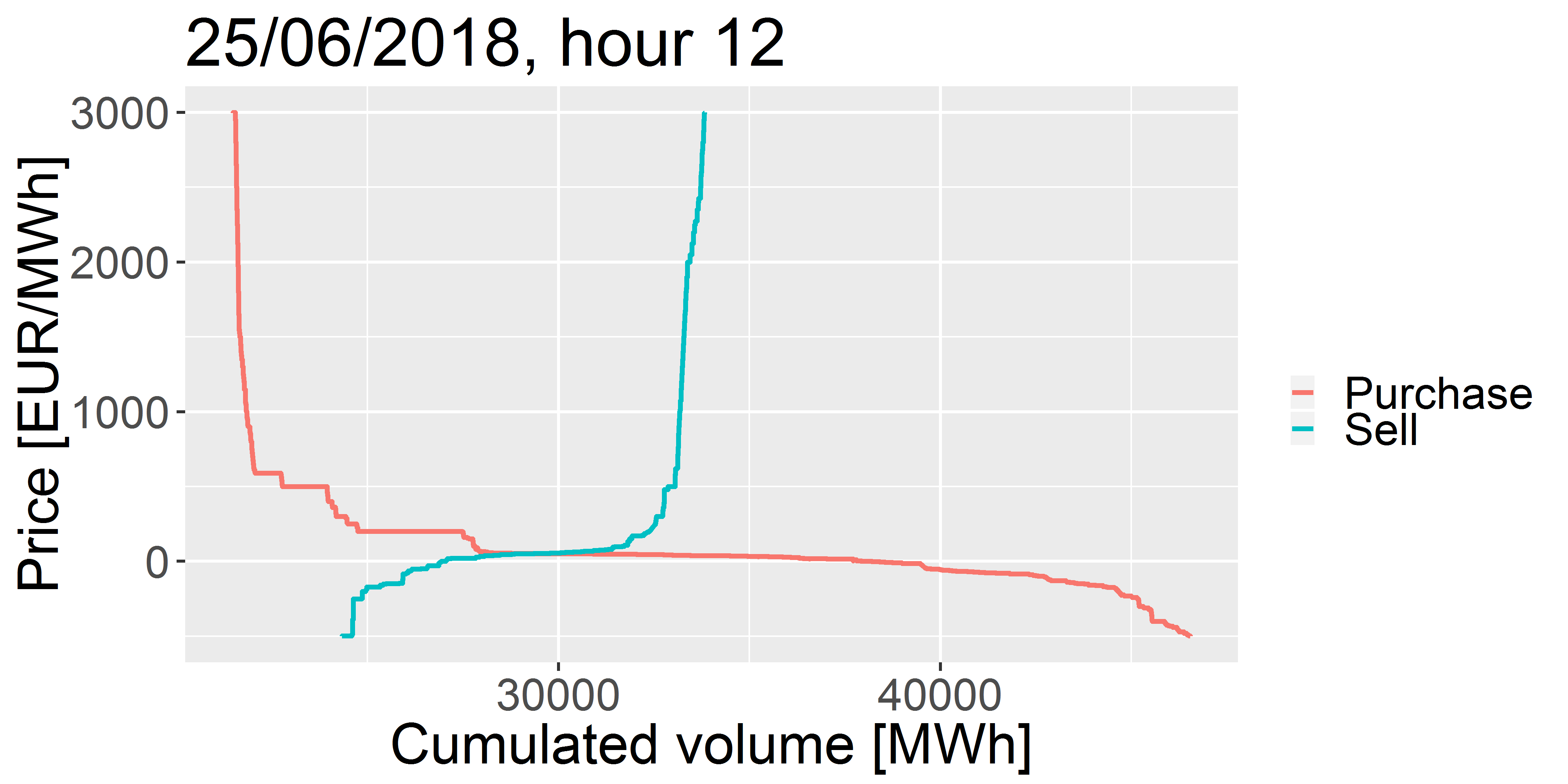}
\includegraphics[width=.49\textwidth]{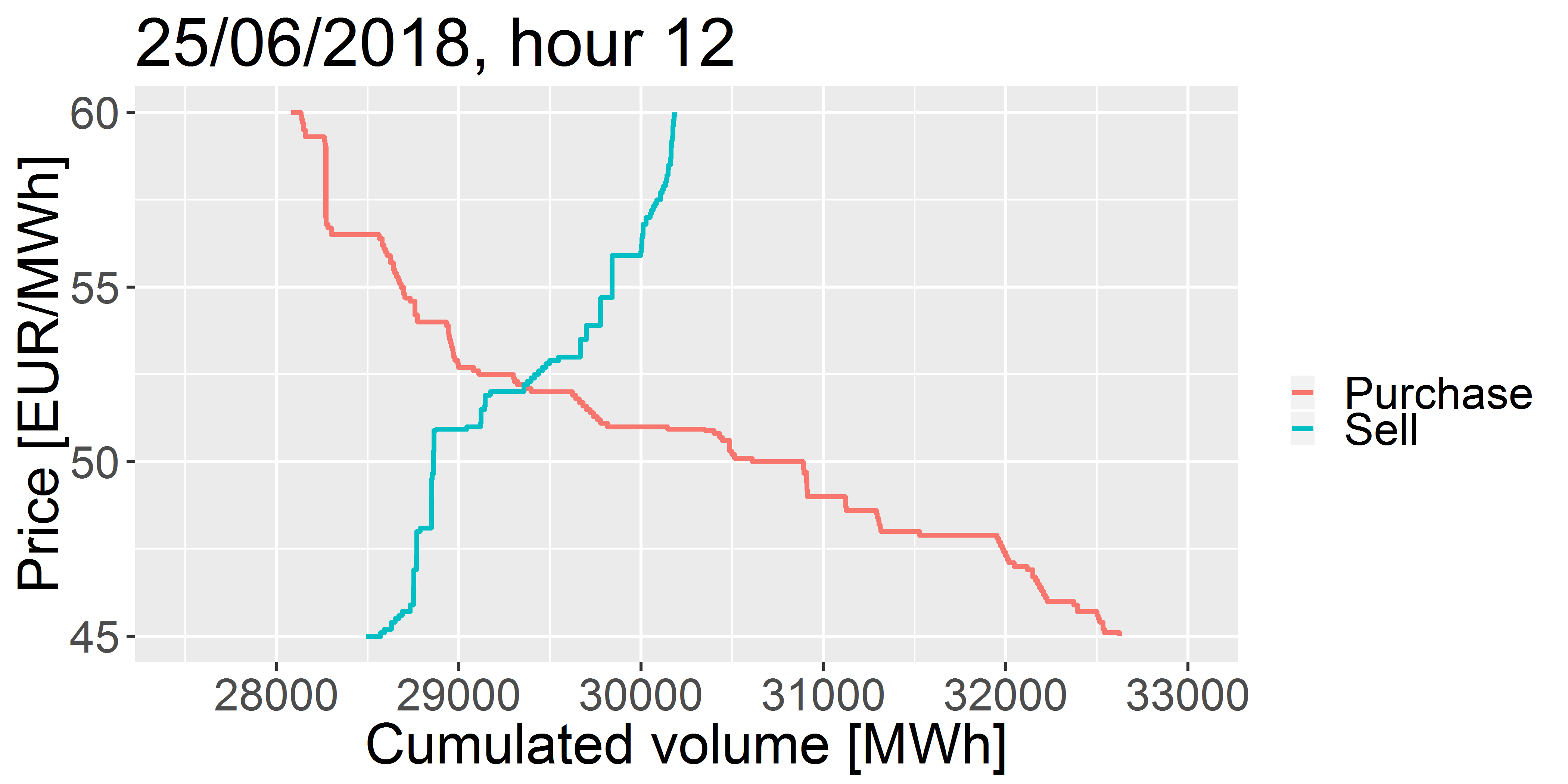}
\caption{Order book data for a particular hour on different scales.}
\label{fig:orderBook}
\end{figure}

In the following, we shortly explain the idea of order-book-based price forecasts.
Each price is the result of an auction, which can be represented as a bid and an ask curve. 
For a particular hour, those curves are shown in Figure~\ref{fig:orderBook}.
The intersection of the bid (purchase, demand) and ask (sell, supply) curve is the market clearing price (MCP).
In the magnified figure, it is clearly visible that the bid and ask curves are step functions.
Each step width is the cumulated volume which market participants have put in the auction at a certain price. 
Price levels correspond in fact to the marginal production costs of different power plants.
Due to the regulatory environment, in particular renewables bid at negative prices in the auction.
Moreover, in contrast to a classical power plant, the produced amount of renewable energy is stochastic and total expected production is sold on the exchange.
Relying on those economical circumstances, the order-book-based forecasting modifies the volumes at different price levels in the bid and ask curves.
The modifications correspond to the forecasted wind and solar power infeed. 
An important issue is \textit{which} price levels are influenced by the renewable infeed.
Usually, energy utilities use exhaustive statistical analysis on historical data to identify the price levels and the impact of the renewable forecasts.
In fact, there are also other fundamental factors which influence the market price, first of all the expected electricity demand.
This paper focuses on machine learning methods to reduce the effort for building a forecast model.

In the following section we give an overview on existing literature on the economics of electricity markets, order-book-based models and the use of machine learning in price forecasting.
In Section~\ref{sec:featureExtraction} we detail our methodology.
Section~\ref{sec:results} is devoted to numerical results and a comparison to other models from the literature.
Section~\ref{sec:conclusions} concludes.

\section{Existing literature on price forecasting and machine learning in electricity markets}
\begin{figure}[tb]
\includegraphics[width=.49\textwidth]{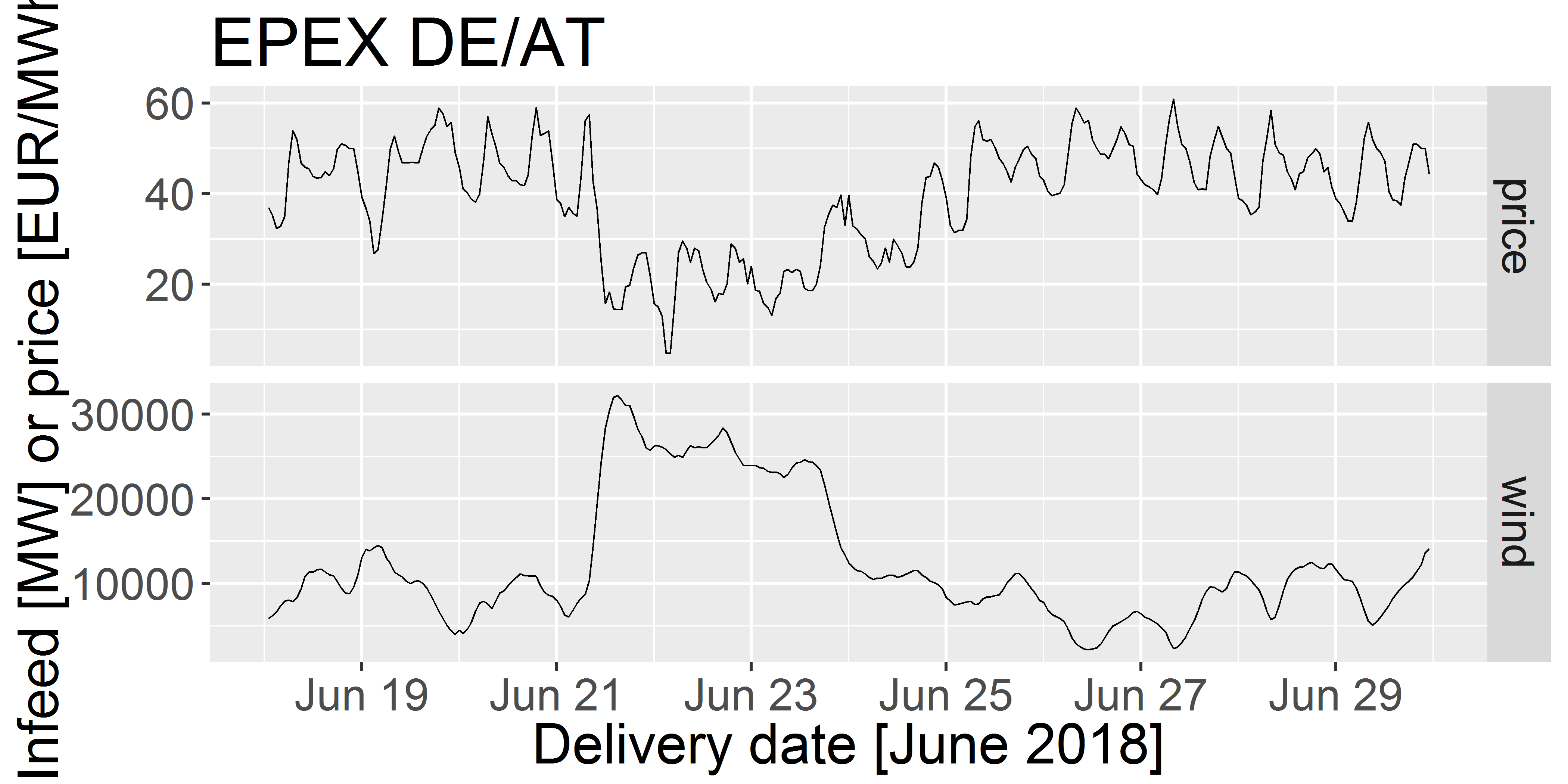}
\includegraphics[width=.49\textwidth]{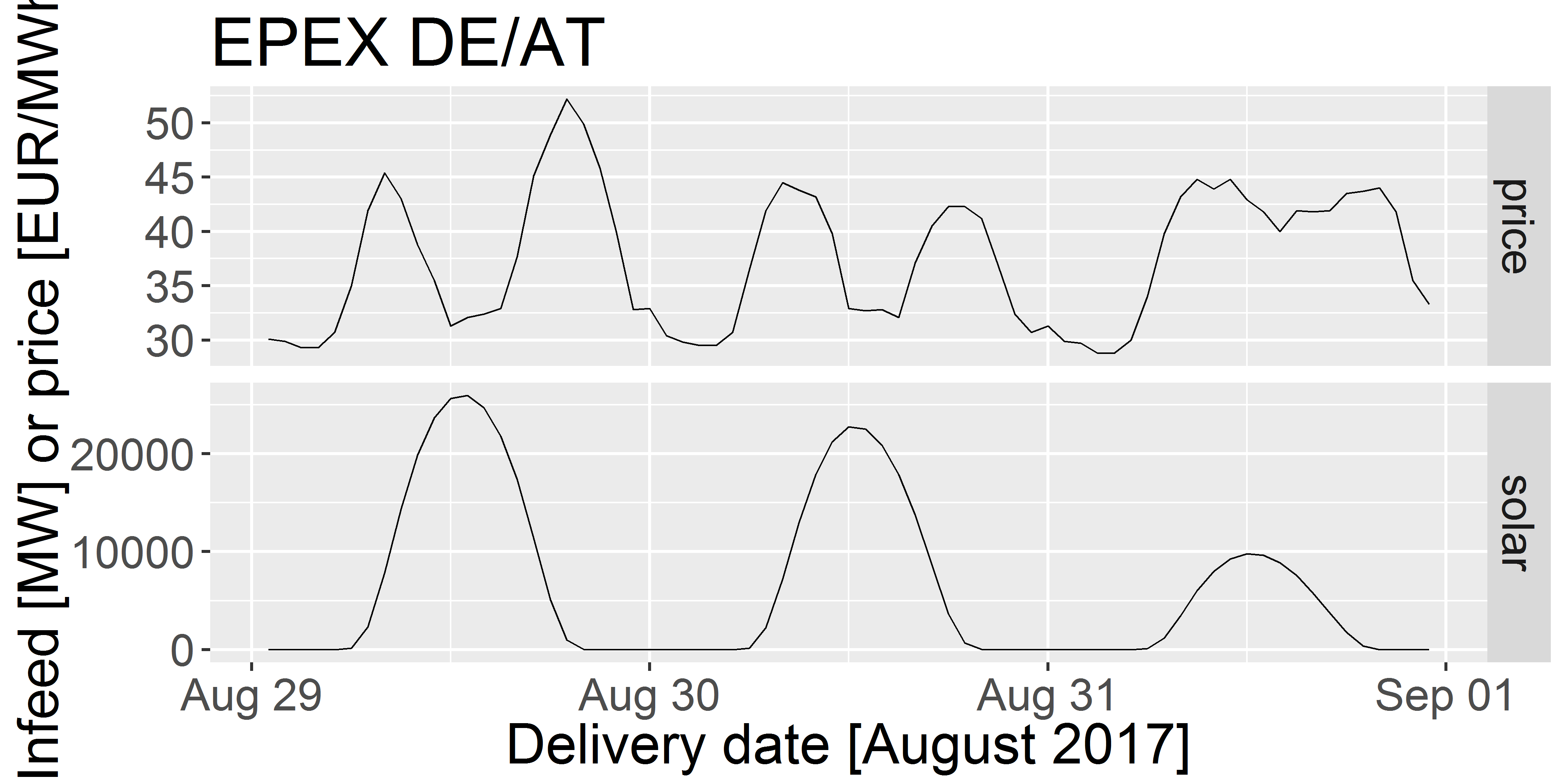}
\caption{Influence of wind and solar infeed on price.}
\label{fig:windsolarinfluence}
\end{figure}

Solar and wind energy is playing a more and more prominent role in today's electricity mar\-kets. Empirical studies show that renewable electricity generation is both highly volatile and has a substantial impact on the day-ahead electricity price (\cite{wagner2014}).
Using multivariate regression methods, various authors have quantified the influence renewable infeed has on the price (\cite{cludius2014,wuerzburg2013}). 
Higher renewable infeed generally leads to lower market prices.
The impact on the hourly prices is demonstrated separately by typical profiles for wind and solar energy in Figure~\ref{fig:windsolarinfluence}.
Therefore, we also use expected solar and wind infeed as features for the price forecasts. 

There is a vast body of literature on electricity price forecasting, over which \cite{aggarwal2009} give an early overview. Their survey covers 47 papers published between 1997 and 2006 with topics ranging from game theoretic to time series and machine learning models. 
A more recent extensive literature overview is given by \cite{weron2014}, in which the author distinguishes and describes five model classes for electricity price forecasting, namely game-theoretic, fundamental, reduced-form, statistical and machine learning models.  
The article closes with a discussion of future challenges in the field, including the issues of feature selection, probabilistic forecasts, combined estimators, model comparability and multivariate factor models.
Regarding this last aspect, \cite{ziel2018} conduct an empirical comparison of different univariate and multivariate model structures for price forecasting. Comparing a total of 58 models on several datasets, they find that there is no single modelling framework that consistently achieves the best results. 

Statistical methods which have been applied to price forecasting include, for example, dynamic regression and transfer functions (\cite{nogales2002}), wavelet transformation followed by an ARIMA model (\cite{conejo2005}) and weighted nearest neighbor techniques (\cite{troncoso2007}). 
There are many applications of machine learning methods in electricity price forecasting.~\cite{amjady2006} compare the performance of a fuzzy neural network with one hidden layer to ARIMA, wavelet-ARIMA, multilayer perceptron and radial basis function network models for the Spanish market.~
\cite{chen2012} also use a neural network with one hidden layer and a special training technique called extreme learning machine on Australian data. On the same market, \cite{mosbah2016} train a multilayer neural network on temperature, total demand, gas price and electricity price data of the year 2005 to predict hourly electricity prices for January 2006.
In order to show the superior performance of neural networks compared to time series approaches, \cite{keles2016} conduct an extensive study focussing on the important topics of variable selection and hyperparameter optimisation. They select the most predictive features via a k-nearest neighbor backward elimination approach and employ 6-fold cross-validation to optimise forecasting performance over several hyperparameters of the neural network. The resulting network is found to outperform the benchmark models substantially.
Recently, more sophisticated types of neural networks have been used: In a benchmark study, \cite{lago2018} compare feed-forward neural networks with up to 2 hidden layers, radial basis function networks, deep belief networks, convolutional neural networks, simple recurrent neural networks, LSTM and GRU networks to several statistical and also to other machine learning methods like random forests and gradient boosting. Using the Diebold-Mariano test, they show the deep feed-forward, GRU and LSTM network approaches to perform significantly better than most of the other methods on Belgium market data.~
\cite{marcjasz2018} consider a non-linear autoregressive (NARX) neural network-type model which especially accounts for the long-term price seasonality. Also using the Diebold-Mariano test, they show that this approach can improve the accuracy of day-ahead forecasts relative to the corresponding ARX benchmark.

Among the features considered in the aforementioned studies historical electricity prices, total demand series, total demand prognoses, renewable infeed forecasts, weather data and calendar information appear on a regular basis.
On the other hand, to the best of our knowledge, the first to use supply and demand curves for price prediction are \cite{ziel2016}. 
Their goal is to fill the gap between time series analysis and structural analysis by setting up a time series model for these curves and then forecasting the future market clearing price as the intersection of the corresponding forecasted curves.
They compare multiple time series prediction methods based on this approach.
However, they do not investigate whether the performance of their model can be enhanced by machine learning techniques.

\section{Methodology} \label{sec:featureExtraction}
\paragraph{Data preparation and feature extraction from order book} 
Our dataset ranges from 1/2/2015 to 18/9/2018 (31823 single auctions) and includes order book data from the EPEX German/Austrian electricity spot market, transparency data from EEX on expected wind and solar power infeed, and expected total demand data from ENTSO-E. 
To avoid data dredging, $20\%$ (about 9~months) of the available data at the end of the time period are held back for an out-of-sample model evaluation (see Section~\ref{sec:results}).

For feature extraction, i.e., translating the order book into a vector of numbers, we use ideas from \cite{coulon2014} and \cite{ziel2016}. 
Let $\mathbb{P} := \{-500, -499.9, \dots, 2999.9, 3000\}$ be the set of possible prices and $\mathbb{T} := \{t_1, t_2, \dots, t_{T-1}, t_T\}$ the set of time points for which there are data available.
Each $t = (d, h)\in\mathbb{T}$ is a tuple consisting of a date $d\in\{\text{1/2/2015}, \text{2/2/2015}, \dots, \text{18/9/2018}\}$ and an hour $h\in\{0, 1, \dots, 23\}$.
We represent the supply and demand data at time $t\in\mathbb{T}$ as vectors $\left(V_t^S(P)\right)_{P\in\mathbb{P}}$ and $\left(V_t^D(P)\right)_{P\in\mathbb{P}}$, where $V_t^S(P)$ and $V_t^D(P)$ denote the supply and demand volume, respectively, bid at price level $P\in\mathbb{P}$. 
The market clearing price at time $t$ is determined by EPEX via the EUPHEMIA algorithm, which also considers complex orders. 
There is no information about such orders in our dataset, so it would be unreasonable to expect any learning algorithm to incorporate them into its price prediction. 
Therefore, we calculate the market clearing price that would result from considering only the available supply and demand data and use this as the target value for price prediction.
To this end, we define the so-called supply and demand curves
\begin{align}
S_t(P) &:= \sum\limits_{p\in\mathbb{P}, \, p\leq P} V_t^S(p) \text{ and} \\
D_t(P) &:= \sum\limits_{p\in\mathbb{P}, \, p\geq P} V_t^D(p).
\end{align} 

The MCP $P_t^*$ lies at the intersection of the supply and demand curves. 
As $S_t$ and $D_t$ are step functions, explicit formulae for $P_t^*$ are quite technical and therefore omitted. 
We refer to Figure~\ref{fig:orderBook} for a graphical illustration.
To reduce the dimensionality, we partition $\mathbb{P}$ into price classes and use the volumes per price class as features.
To determine these classes we use a heuristic which aims to achieve that they all contain approximatively the same amount of volume on average.
This ensures that there are more price classes at the interesting parts of the curve, i.e., in the price regions with many bids.
We begin by averaging the supply curves over all time points, obtaining
\begin{align}
\overline{S}(P) := \frac{1}{T} \sum\limits_{t\in\mathbb{T}} S_t(P) \text{ for } P\in\mathbb{P}.  
\end{align}
Then, we fix a volume~$V_* > 0$ that each price class is supposed to contain on average and partition $\mathbb{P}$ accordingly.
We compute the number of classes,
\begin{align}
\tilde{M}^S := \min\left\{i\in\mathbb{N}: iV_* > \overline{S}(3000) \right\}, % \text{ and} \\
% \tilde{M}^D := \min\left\{j\in\mathbb{N}: jV_* > \overline{D}(-500)\right\},
\end{align}
and set
\begin{equation}
\tilde{c}_i^S := \begin{cases}
\min \left\{P\in\mathbb{P}: \overline{S}(P) \geq iV_* \right\} & \text{for } i = 0, \dots, \tilde{M}^S - 1 \\
3000 & \text{for } i = \tilde{M}^S
\end{cases}.
\end{equation}
Now, we define the classes 
\begin{align} \label{eq:priceclasses} 
\tilde{\mathbb{C}}_i^S := \begin{cases}
[\tilde{c}_{i-1}^S, \tilde{c}_i^S] & \text{for } i = 1 \\
(\tilde{c}_{i-1}^S, \tilde{c}_i^S] & \text{for } i = 2, \dots, \tilde{M}^S 
\end{cases}, 
\end{align}
where we take $(c, c]$ or $[c, c]$ to denote the singleton $\{c\}$.
By removing intervals that occur multiple times, we get pairwise different price classes $\mathbb{C}_i^S = (c_{i-1}^S, c_i^S]$ as well as corresponding volume features 
\begin{align} \label{eq:pricesonclasses}
V_t^S(\mathbb{C}_i^S) &:= \sum\limits_{P\in\mathbb{C}_i^S} V_t^S(P) \text{ for } i\in\{1, \dots, M^S\},
\end{align}
where $M^S \leq \tilde{M}^S$ denotes the numbers of \textit{unique} price classes.
\begin{figure}[tb]
	\centering
		\includegraphics[scale = 0.7]{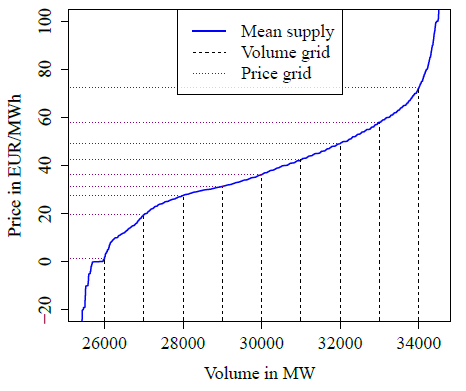}
		\caption{Averaged supply curve with $V_* = 1000$, figure taken from \cite{ziel2016}.}
\label{fig:priceclasses}
\end{figure}
While the details are rather technical (see also \cite{ziel2016}), the graphical illustration in Figure~\ref{fig:priceclasses} should make the idea intuitively clear.
Following the same logic and using analogous notation, we obtain boundaries $c_0^D, c_1^D, \dots, c_{M^D}^D$ and corresponding price classes for the demand curves.
Similarly as with the original supply and demand curves, one can calculate the price that results from the price classes and of course, in general, does not exactly coincide with the actual market clearing price~$P_t^*$.

\begin{figure}[tb]
\includegraphics[width=\textwidth]{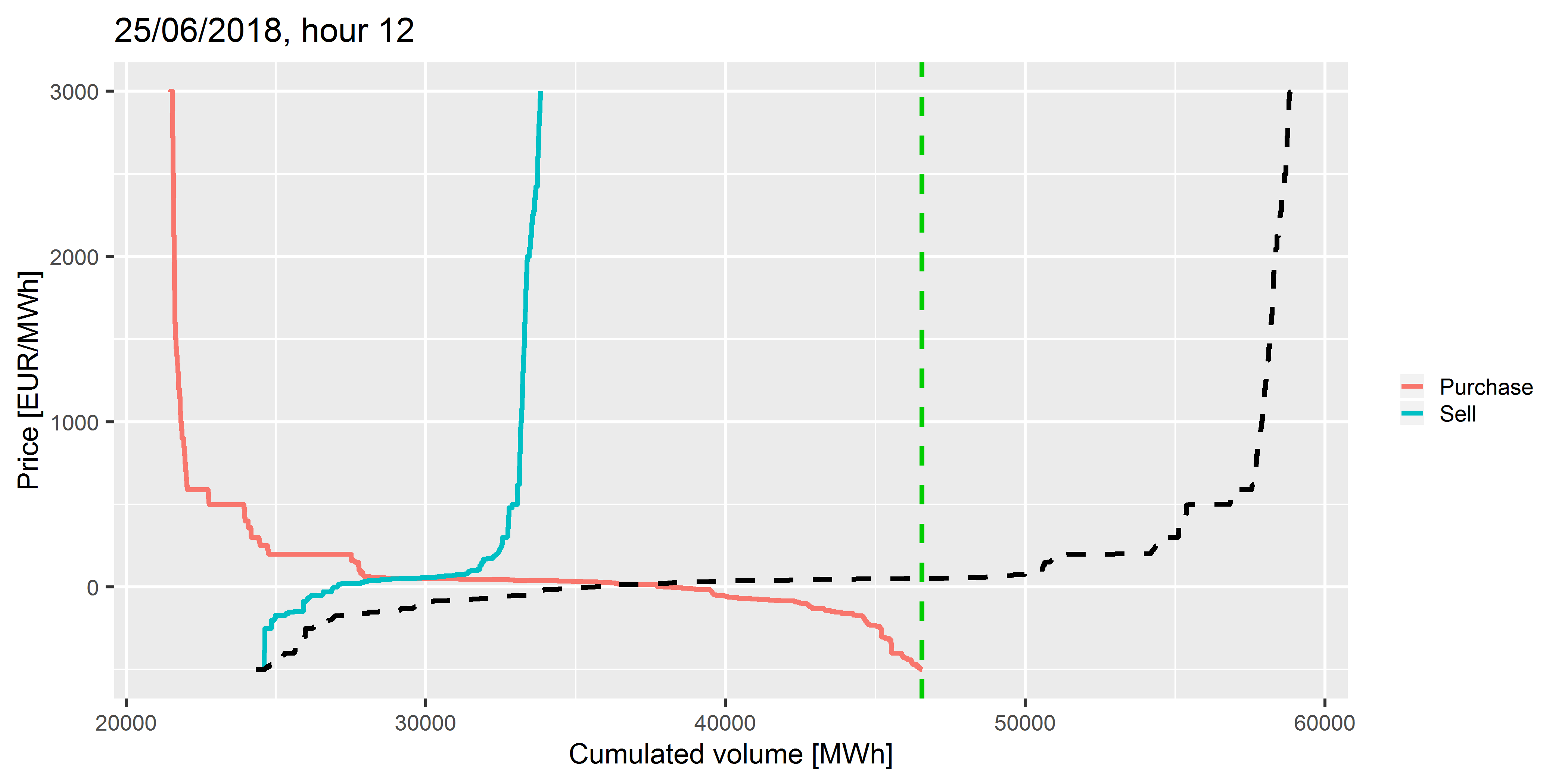}
\caption{Transformed supply and demand curves to price curve (dashed black) and inelastic demand (dashed green vertical line).}
\label{fig:pricecurve}
\end{figure}

Finally, in order to simplify both implementation and interpretation without losing any essential information, we transform the supply and demand features into a so-called price curve. 
For this, let $\{c_0^X, c_1^X, \dots, c_{M^X}^X\}$ be the ascendingly ordered union of the supply and demand price class boundaries. 
Proceeding from this, similarly to~(\ref{eq:priceclasses}) and~(\ref{eq:pricesonclasses}), we define new price classes $\mathbb{C}_1^X, \mathbb{C}_2^X, \dots, \mathbb{C}_{M^X}^X$ 
and volume features
\begin{equation}
V_t^X\left(\mathbb{C}_k^X\right) := \sum\limits_{P\in\mathbb{C}_k^X} \left(V_t^S(P) + V_t^D(P)\right).
\end{equation}
We use these price curve features and additionally the total demand $D_t(-500)$ as inputs for the price prediction.
Figure~\ref{fig:pricecurve} shows an example of such a price curve calculated from given supply and demand curves.

There is also an economic interpretation for this transformation:
In fact, electricity demand is highly price-inelastic, so the constant inelastic demand is the expected total demand for electricity at that hour.
The price curve is the so-called merit order, which represents the electricity production units sorted by their variable production costs.
For more details, we refer to standard literature on electricity markets like \cite{burger2014}.
Note that the price curve still contains the information that is necessary to calculate the resulting price: 
The MCP lies at the intersection of the cumulative price curve and the constant inelastic demand.
In addition to the price curve and inelastic demand, we use as features the renewable infeed and total demand forecasts as well as some calendar information, namely
\begin{itemize}
\item year as a numerical variable, 
\item a binary variable on daylight saving time,
\item type of day as a one-hot-encoded categorical variable with three different values (workday, Saturday/bridge day, Sunday/holiday), 
\item month, and 
\item hour. 
\end{itemize}
To account for the periodicity of months and hours, we project these values on a circle and use the two-dimensional projections as features.
More precisely, if date $t\in\mathbb{T}$ lies in month $m\in\{1, \dots, 12\}$ and refers to hour $h\in\{0,\dots,23\}$, this is encoded as
\begin{align}
\text{month}_x(t) &:= \sin\left(\frac{2\pi m}{12}\right), \,
\text{month}_y(t) := \cos\left(\frac{2\pi m}{12}\right) \text{ and } \\
\text{hour}_x(t)  &:= \sin\left(\frac{2\pi h}{24}\right), \,
\text{hour}_y(t)  := \cos\left(\frac{2\pi h}{24}\right).
\end{align}

For prediction, we use the price curve features of a preceding day, the so-called reference date $r(d)$. 
For notational convenience, we write $r(t) := (r(d), h)$.
As a reference date for $d$ we use the nearest day before $d$ which is of the same type of day as $d$. This is a simple but efficient technique in energy economics.
More sophisticated methods to define a reference date may incorporate similarities in renewable infeed and demand profile.

\paragraph{Training of learning algorithms} \label{sec:algorithms}

We employ ordinary linear regression, random forests and feed-forward neural networks to predict hourly electricity prices. 
Note that we use the prices which are implied by the volume features $V_t^X(\mathbb{C}_k^X)$ as target values, which means that the prices we aim to forecast attain the values $c_0^X, c_1^X, \dots, c_{M^X}^X$.  
On the whole dataset, the absolute difference between these price approximations and the real prices is $2.07$ EUR/MWh on average (corresponding to a median absolute percentage deviation of $4.28\%$). 
While we assume ordinary linear regression to be well-known, we give a brief description of the machine learning algorithms we consider.
In each case, our goal is to approximate the function~$f:\mathbb{R}^N\to\mathbb{R}$ which maps the features described above to the corresponding electricity price. 
To this end, we assume to be given a set of training data $\{(x_1, y_1), \dots, (x_n, y_n)\}$ where 
\begin{equation}
y_i = f(x_i) + \varepsilon_i, x_i\in\mathbb{R}^N, i = 1, \dots, n,
\end{equation} 
and $\left(\varepsilon_1, \dots, \varepsilon_n\right)$ is a vector of realizations of independent, homoskedastic random variables with zero expectation.

\paragraph{Random forests}
Random forests are based on a simpler machine learning method called decision trees (\cite[chapter 9.2]{hastie2001}).
While decision trees are easy to understand, they often perform rather poorly because of their high dependence on the training data. 
Random forests aim to overcome this drawback by averaging the predictions of several decision trees that are trained in a randomized way proceeding from the same data (\cite{breiman2001}). 
As part of their training process, random forests offer a convenient way to assess the influence of each feature on the output. 
Therefore, they can deliver a ranking of the features  according to their relevance for electricity price prediction.
While it is quite interesting in its own right, we also use this ranking for feature selection, i.e., for training a feed-forward neural network only on the $N_F\in\mathbb{N}$ most important features (e.g. $N_F = 10$).

\paragraph{Feed-forward neural networks} 
Feed-forward neural networks can be viewed as a far-reaching non-linear extension to ordinary linear regression. 
They consist of several \textit{layers}, through which the input is fed via the composition of non-linear activation functions and weighted sums in order to generate the output.  
The smallest unit (one vector component) of such a layer is called a neuron.
A central result in the theory of neural networks states that, using a non-constant, bounded and continuous activation function, a neural network with just one hidden layer can in principle approximate any continuous function arbitrarily well when there are sufficiently many neurons and appropriate weights are chosen (\cite{hornik1991}).
In practice, a higher number of layers has been found to improve performance for many applications (\textit{deep learning}).
Besides the number of hidden layers and the number of neurons per layer, there are other so-called \textit{hyperparameters} on which forecasting performance can critically depend. 
For instance, the optimisation algorithm that is used to train the network has to be chosen. Typically, some variant of stochastic gradient descent (SGD) like rmsprop (\cite{tieleman2012}) or Adam (\cite{kingma2015}) is used.
Furthermore, SGD-type algorithms work with batches of training data.
The \textit{batch size} can be varied in order to improve performance.
Other hyperparameters which we consider include the number of epochs, i.e., the number of times the training data are fed into the optimisation algorithm, the activation function (tangens hyperbolicus, rectified linear unit, identity) and whether to employ dropout to avoid overfitting (\cite{srivastava2014}) and batch normalization to avoid internal covariate shift (\cite{ioffe2015}).

\paragraph{Hyperparameter optimisation via cross-validation}
We choose the hyperparameter values for the neural networks and random forests using five-fold cross-validation.
First, we define a grid of hyperparameter combinations to be evaluated. 
Then, for every combination of hyperparameters in the grid, we split our training dataset into five parts or \textit{folds} of equal size, train a model with these values on four of the folds and evaluate its performance on the remaining one. 
After repeating this five times, each time with a different validation fold, we average performances. 
Finally, once the whole grid has been evaluated, we choose the hyperparameter combination that performs best on average. 

Summarising, the features we use to forecast the spot price of a time point $t$ with reference time point $r(t)$ are
\begin{itemize}
\item the total demand $D_{r(t)}(-500)$ and the price curve features of the same hour on the reference day, i.e., $V_{r(t)}^X(\mathbb{C}_k^X)$, $k=1, \dots, M^X$,
\item the solar and wind infeed forecasts as well as the total demand forecast for the time points $t$ and $r(t)$,
\item the calendar features year, daylight saving time, type of day, month and hour for the time points $t$ and $r(t)$.
\end{itemize}

We considered about 100 different parameter combinations for the random forests with the number of trees equal to $10$, $100$, $1000$, $5000$, $10000$ or $50000$.
For the neural networks, we tested over 1000 parameter combinations with about 20 different network sizes ranging from one hidden layer with 5 neurons to $100$ hidden layers with 25 neurons each (see Table~\ref{tab:hyperparameters}). 

\section{Results} \label{sec:results}

% Error measures
To evaluate model performance, we primarily use the root-mean-square error
\begin{equation*}
\text{RMSE} :=\left(\frac{1}{n} \sum\limits_{i=1}^n \left(\hat{y}_i-y_i\right)^2 \right)^{1/2},
\end{equation*}
where $\hat{y}_i$ are the predictions, $y_i$ are the true target values and $n$ is the number of observations for which a prediction is made. 
Furthermore, we consider the mean absolute error 
\begin{equation*}
\text{MAE} := \frac{1}{n} \sum\limits_{i=1}^n \left| \hat{y}_i - y_i \right|
\end{equation*}
as a more interpretable measure of how far off the prediction is on average.
The RMSE is the error measure which the machine learning algorithms aim to minimize during training. 
Accordingly, we select the model architecture that performs best in the 5-fold cross-validation with respect to the RMSE. 
In the electricity forecasting literature, sometimes the mean absolute percentage error (MAPE) is used.
This is unsuitable for the German market, as often the MCP is at or close to zero.
Therefore, we report the median absolute percentage error 
\begin{equation*}
\text{MdAPE} := \med \left\{ \frac{| \hat{y}_i - y_i |}{| y_i |}, i=1, \ldots, n  \right\}
\end{equation*}
for comparison.

% Naive Benchmark
Aside from the methods which were described in Section \ref{sec:algorithms}, we consider two benchmarks.
The first one is called the naive benchmark (\cite{nogales2002}). 
Its forecast for hour $h$ of date $d$ is the price at hour $h$ of the previous day if $d$ is a workday other than Monday and the price at hour $h$ of the same type of day in the previous week otherwise. 

% EXAA
The second benchmark is based on a different market, the Energy Exchange Austria (EXAA), where the electricity price is fixed two hours before the EPEX auction takes place. 
Therefore, the EXAA price at a time point $t$ can directly be used as a predictor for the EPEX price at the same time point.
In fact, \cite{ziel2015} show this benchmark to be highly competitive. 
However, note that it is not really appropriate to compare the remaining forecasting methods to the EXAA benchmark because they are based on different information (see also \cite{ziel2016}). 
Nonetheless, the EXAA benchmark can provide some orientation on how well other models perform and how much improvement could be expected. 

% Random forest
The best-performing random forest consists of 1000 decision trees where at each step in the training of the underlying decision trees a randomly chosen subset of size 23 (corresponding to $25\%$) of all available features is used and where a tree node is only split further if it contains at least $1\%$ of all training data. 
We also use the random forest to support feature selection for the following neural network approach.

% Feed-forward neural network
For the neural networks under consideration we use different feature vector realizations: 
\begin{itemize} 
\item all available features, 
\item all but the price curve features of the reference date,
\item the 10 most influential features according to the best-performing random forest,
\item the 20 most influential features according to the best-performing random forest.
\end{itemize} 
For each case we use different network architectures determined by hyperparameter optimisation as described above.
These are reported in Table~\ref{tab:hyperparameters} where each column corresponds to a choice of features and each row corresponds to a hyperparameter. 
The notation $[5, 5, 5]$ for the network architecture means that a 3-layer network consisting of 5 nodes per layer is used.
For the networks that are trained on the selected features we find a deeper architecture to perform best: $[25]$~*~$25$ denotes a $25$-layer network with $25$ nodes per layer.
Analogously, in the dropout row, $[0, 0.25, 0]$ means that dropout is employed with a probability of $25\%$ after the second layer and $[0.1]$~*~$25$ means that dropout is employed after each of the $25$ layers with a probability of $10\%$.
It is noteworthy that the best-performing network when using all features is rather small. 
Thus, as an additional plausibility check, we also consider the network architectures proposed by \cite{keles2016} (network size $[48, 48]$, sigmoid activation, no dropout) and \cite{lago2018} (network size $[239, 162]$, relu activation, no dropout). Note that their models do not consider price curve features, i.e., order book data.

\begin{table}[ht]
\caption{The hyperparameters which were used when training feed-forward neural networks with different features (all, without curve features, only with the 10 or 20 most influential features as chosen by the best-performing random forest).}
\centering
\begin{tabular}{|l|c|c|c|c|}
\hline
Hyperparameter & \multicolumn{1}{l|}{All features} & \multicolumn{1}{l|}{\begin{tabular}[c]{@{}l@{}}Without curve \\ features\end{tabular}} & \multicolumn{1}{l|}{\begin{tabular}[c]{@{}l@{}}Selected features \\ ($N_F = 10$)\end{tabular}} & \multicolumn{1}{l|}{\begin{tabular}[c]{@{}l@{}}Selected features \\ ($N_F = 20$)\end{tabular}} \\ \hline
\begin{tabular}[c]{@{}l@{}}Network \\ architecture\end{tabular} & [5, 5, 5] & [5, 5] & [25] * 25 & [25] * 25   \\ \hline
Optimiser & rmsprop & Adam & Adam & Adam \\ \hline
\begin{tabular}[c]{@{}l@{}}Number of \\ epochs\end{tabular} & 100 & 100 & 100 & 100 \\ \hline
Batch size & 128 & 64 & 128 & 128 \\ \hline
\begin{tabular}[c]{@{}l@{}}Activation \\ function\end{tabular} & tanh & relu & relu & relu  \\ \hline
Dropout & [0, 0.25, 0] & [0, 0.25, 0] & [0.1] * 25 & [0.1] * 25  \\ \hline
\begin{tabular}[c]{@{}l@{}}Batch \\ normalization\end{tabular} & no & yes & yes & yes \\ \hline
\end{tabular}
\label{tab:hyperparameters}
\end{table}

The results of the chosen model configurations are shown in Table~\ref{tab:results}. The errors we report are measured both on the training set (in-sample error) to evaluate how well the model describes the given data and on the test set (out-of-sample error) to assess model performance on previously unseen data (20\% of our whole dataset).

\begin{table}[ht]
\caption{Comparison of in-sample and out-of-sample errors in EUR per MWh or $\%$ for various price forecasting techniques.}

\centering
\begin{tabular}{|l|l|l|l|l|l|l|}
\hline
\multirow{2}{*}{Forecasting technique} & \multicolumn{3}{l|}{in-sample error} & \multicolumn{3}{l|}{out-of-sample error} \\ \cline{2-7}
 & RMSE & MAE & MdAPE & RMSE & MAE & MdAPE \\ \hline
Naive model & 13.55 & 7.87 & 15.31\% & 12.68 & 7.71 & \bf{11.61\%} \\ \hline
Ordinary linear regression & 6.85 & 4.25 & 10.93\% &  9.60 & 7.52 & 16.95\% \\ \hline
Random forest & 6.77 & 4.17 & 9.73\% & 11.92 & 9.32 & 19.9\%  \\ \hline
\begin{tabular}[c]{@{}l@{}}Feed-forward neural network\\ with architecture from \cite{keles2016}\end{tabular} & 6.72 & 4.51 & 11.49\% & 14.87 & 12.81 & 30.63\%  \\ \hline
\begin{tabular}[c]{@{}l@{}}Feed-forward neural network\\ with architecture from \cite{lago2018}\end{tabular} & 2.27 & 1.65 & 4.45\% & 21.05 & 8.94 & 15.22\%  \\ \hline
Feed-forward neural network & 5.45 & 3.57 & 8.89\% &  9.59 & \bf{7.08} & 14.18\% \\ \hline
\begin{tabular}[c]{@{}l@{}}Feed-forward neural network\\ without curve features\end{tabular} & 6.63 & 4.41 & 11.22\% & 10.11 & 7.85 & 16.12\% \\ \hline
\begin{tabular}[c]{@{}l@{}}Feed-forward neural network \\ with feature selection ($N_F = 10$) \end{tabular} & 7.69 & 5.06 & 11.68\% & \bf{9.41} & 7.34 & 15.57\% \\ \hline
\begin{tabular}[c]{@{}l@{}}Feed-forward neural network \\ with feature selection ($N_F = 20$) \end{tabular} & 7.71 & 4.95 & 11.27\% & 13.65 & 10.18 & 21.48\% \\ \hline \hline
EXAA & 6.47 & 3.53 & 7.56\% & 5.58 & 3.92 & 7.22\% \\ \hline
\end{tabular}
\label{tab:results}
\end{table}

\paragraph{Alternative: More sophisticated network architectures}
Apart from feed-forward neural networks we also analysed recurrent neural networks. 
As electricity spot prices can be expected to depend on previous days' features and prices, it seems reasonable to model them as a multivariate time series. 
While classical approaches like ARIMA or GARCH models are possible, this also is a typical application for recurrent neural networks because they explicitly incorporate the sequential structure of the inputs. 
In this case, the goal was to predict the $24$-dimensional vector of spot prices at some date $d$ based on information available up to date $d-1$. 
For each date $e\leq d-1$ this information consists of the curve features for date $e$ as well as the calendar features, expected renewable infeed and total demand for date $e+1$.
We implemented this approach using the \textit{long short-term memory} (LSTM) architecture that allows for efficient training of recurrent neural networks (\cite{hochreiter1997}), but the results were not as convincing as with the other methods. 
This might be due to the high dimensionality of the multivariate time series under consideration.
Therefore, we focused on the random forest and feed-forward neural network approaches where the temporal dependence structure is more explicitly incorporated as a feature by means of the reference day.

\section{Conclusion}\label{sec:conclusions}
Our results show that neural networks can indeed provide order-book-based price forecasts with competitive results.
However, they do not perform significantly better than simpler methods like ordinary linear regression.
Whereas the \textit{classical} order-book-based forecasting technique requires a lot of statistical analysis, the network architecture optimisation also demands significant resources.
We also found that reducing the number of features generally improves results.
In regard to the RMSE, we find that the feed-forward neural network with only 10 features as selected by the random forest performs best. 
Considering the MAE (a measure directly linked to revenues from financial trading), the feed-forward neural network without feature selection is in the lead.
However, the naive model shows good results as well, supporting this traditional and often applied heuristic in energy economics.
The neural network architectures from literature show competitive in-sample results, but their performance drops significantly in an out-of-sample analysis. 
This indicates overfitting.

The posed research questions have been answered. 
We have shown how to incorporate order book features using volume-based partitioning, a transformation to price curves and feature selection based on random forests.
We have also shown that machine learning cannot significantly reduce the work effort needed in the model set-up, but gives competitive results.

The models do have a lot of potential for improvement.
For instance, there are much more accurate wind and solar infeed forecasts available in the market compared to the data from EEX transparency (unfortunately they are not free of charge).
We see the largest potential in a daily recalibration of the models including an updated feature selection which allows the model to react to fundamental changes in the market (coal and gas prices, power plant outages, ...).

In addition, we also analysed different applications of machine learning on EPEX order books, which are not outlined in detail:
We employed neural networks to reconstruct renewable infeed from the order book and used the networks to generate price forward curves.

\FloatBarrier
\bibliographystyle{apacite}
\bibliography{literature}

\begin{thebibliography}{}

\bibitem [\protect \citeauthoryear {%
Aggarwal%
, Saini%
\BCBL {}\ \BBA {} Kumar%
}{%
Aggarwal%
\ \protect \BOthers {.}}{%
{\protect \APACyear {2009}}%
}]{%
aggarwal2009}
\APACinsertmetastar {%
aggarwal2009}%
\begin{APACrefauthors}%
Aggarwal, S\BPBI K.%
, Saini, L\BPBI M.%
\BCBL {}\ \BBA {} Kumar, A.%
\end{APACrefauthors}%
\unskip\
\newblock
\APACrefYearMonthDay{2009}{}{}.
\newblock
{\BBOQ}\APACrefatitle {{Electricity price forecasting in deregulated markets: A
  review and evaluation}} {{Electricity price forecasting in deregulated
  markets: A review and evaluation}}.{\BBCQ}
\newblock
\APACjournalVolNumPages{International Journal of Electrical Power \& Energy
  Systems}{31}{1}{13--22}.
\newblock
\begin{APACrefDOI} \doi{10.1016/j.ijepes.2008.09.003} \end{APACrefDOI}
\PrintBackRefs{\CurrentBib}

\bibitem [\protect \citeauthoryear {%
Amjady%
}{%
Amjady%
}{%
{\protect \APACyear {2006}}%
}]{%
amjady2006}
\APACinsertmetastar {%
amjady2006}%
\begin{APACrefauthors}%
Amjady, N.%
\end{APACrefauthors}%
\unskip\
\newblock
\APACrefYearMonthDay{2006}{06}{}.
\newblock
{\BBOQ}\APACrefatitle {{Day-Ahead Price Forecasting of Electricity Markets by a
  New Fuzzy Neural Network}} {{Day-Ahead Price Forecasting of Electricity
  Markets by a New Fuzzy Neural Network}}.{\BBCQ}
\newblock
\APACjournalVolNumPages{IEEE Transactions on Power Systems}{21}{}{887--896}.
\newblock
\begin{APACrefDOI} \doi{10.1109/TPWRS.2006.873409} \end{APACrefDOI}
\PrintBackRefs{\CurrentBib}

\bibitem [\protect \citeauthoryear {%
Breiman%
}{%
Breiman%
}{%
{\protect \APACyear {2001}}%
}]{%
breiman2001}
\APACinsertmetastar {%
breiman2001}%
\begin{APACrefauthors}%
Breiman, L.%
\end{APACrefauthors}%
\unskip\
\newblock
\APACrefYearMonthDay{2001}{}{}.
\newblock
{\BBOQ}\APACrefatitle {{Random Forests}} {{Random Forests}}.{\BBCQ}
\newblock
\APACjournalVolNumPages{{Machine Learning}}{45}{}{5--32}.
\newblock
\begin{APACrefDOI} \doi{10.1023/A:1010933404324} \end{APACrefDOI}
\PrintBackRefs{\CurrentBib}

\bibitem [\protect \citeauthoryear {%
Burger%
, Graeber%
\BCBL {}\ \BBA {} Schindlmayr%
}{%
Burger%
\ \protect \BOthers {.}}{%
{\protect \APACyear {2014}}%
}]{%
burger2014}
\APACinsertmetastar {%
burger2014}%
\begin{APACrefauthors}%
Burger, M.%
, Graeber, B.%
\BCBL {}\ \BBA {} Schindlmayr, G.%
\end{APACrefauthors}%
\unskip\
\newblock
\APACrefYear{2014}.
\newblock
\APACrefbtitle {{Managing Energy Risk: An Integrated View on Power and Other
  Energy Markets}} {{Managing Energy Risk: An Integrated View on Power and
  Other Energy Markets}}.
\newblock
\APACaddressPublisher{}{Wiley Finance Series}.
\newblock
\begin{APACrefDOI} \doi{10.1002/9781118618509} \end{APACrefDOI}
\PrintBackRefs{\CurrentBib}

\bibitem [\protect \citeauthoryear {%
Chen%
\ \protect \BOthers {.}}{%
Chen%
\ \protect \BOthers {.}}{%
{\protect \APACyear {2012}}%
}]{%
chen2012}
\APACinsertmetastar {%
chen2012}%
\begin{APACrefauthors}%
Chen, X.%
, Dong, Z.%
, Meng, K.%
, Xu, Y.%
, Wong, K.%
\BCBL {}\ \BBA {} Ngan, H.%
\end{APACrefauthors}%
\unskip\
\newblock
\APACrefYearMonthDay{2012}{11}{}.
\newblock
{\BBOQ}\APACrefatitle {{Electricity Price Forecasting With Extreme Learning
  Machine and Bootstrapping}} {{Electricity Price Forecasting With Extreme
  Learning Machine and Bootstrapping}}.{\BBCQ}
\newblock
\APACjournalVolNumPages{IEEE Transactions on Power Systems}{27}{}{2055--2062}.
\newblock
\begin{APACrefDOI} \doi{10.1109/TPWRS.2012.2190627} \end{APACrefDOI}
\PrintBackRefs{\CurrentBib}

\bibitem [\protect \citeauthoryear {%
Cludius%
, Hermann%
, Matthes%
\BCBL {}\ \BBA {} Graichen%
}{%
Cludius%
\ \protect \BOthers {.}}{%
{\protect \APACyear {2014}}%
}]{%
cludius2014}
\APACinsertmetastar {%
cludius2014}%
\begin{APACrefauthors}%
Cludius, J.%
, Hermann, H.%
, Matthes, F\BPBI C.%
\BCBL {}\ \BBA {} Graichen, V.%
\end{APACrefauthors}%
\unskip\
\newblock
\APACrefYearMonthDay{2014}{}{}.
\newblock
{\BBOQ}\APACrefatitle {{The merit order effect of wind and photovoltaic
  electricity generation in Germany 2008–2016: Estimation and distributional
  implications}} {{The merit order effect of wind and photovoltaic electricity
  generation in Germany 2008–2016: Estimation and distributional
  implications}}.{\BBCQ}
\newblock
\APACjournalVolNumPages{Energy Economics}{44}{}{302--313}.
\newblock
\begin{APACrefDOI} \doi{10.1016/j.eneco.2014.04.020} \end{APACrefDOI}
\PrintBackRefs{\CurrentBib}

\bibitem [\protect \citeauthoryear {%
{Conejo}%
, {Plazas}%
, {Espinola}%
\BCBL {}\ \BBA {} {Molina}%
}{%
{Conejo}%
\ \protect \BOthers {.}}{%
{\protect \APACyear {2005}}%
}]{%
conejo2005}
\APACinsertmetastar {%
conejo2005}%
\begin{APACrefauthors}%
{Conejo}, A\BPBI J.%
, {Plazas}, M\BPBI A.%
, {Espinola}, R.%
\BCBL {}\ \BBA {} {Molina}, A\BPBI B.%
\end{APACrefauthors}%
\unskip\
\newblock
\APACrefYearMonthDay{2005}{May}{}.
\newblock
{\BBOQ}\APACrefatitle {{Day-ahead electricity price forecasting using the
  wavelet transform and ARIMA models}} {{Day-ahead electricity price
  forecasting using the wavelet transform and ARIMA models}}.{\BBCQ}
\newblock
\APACjournalVolNumPages{IEEE Transactions on Power Systems}{20}{2}{1035--1042}.
\newblock
\begin{APACrefDOI} \doi{10.1109/TPWRS.2005.846054} \end{APACrefDOI}
\PrintBackRefs{\CurrentBib}

\bibitem [\protect \citeauthoryear {%
Coulon%
, Jacobsson%
\BCBL {}\ \BBA {} Ströjby%
}{%
Coulon%
\ \protect \BOthers {.}}{%
{\protect \APACyear {2014}}%
}]{%
coulon2014}
\APACinsertmetastar {%
coulon2014}%
\begin{APACrefauthors}%
Coulon, M.%
, Jacobsson, C.%
\BCBL {}\ \BBA {} Ströjby, J.%
\end{APACrefauthors}%
\unskip\
\newblock
\APACrefYearMonthDay{2014}{01}{}.
\newblock
{\BBOQ}\APACrefatitle {{Hourly Resolution Forward Curves for Power: Statistical
  Modeling Meets Market Fundamentals}} {{Hourly Resolution Forward Curves for
  Power: Statistical Modeling Meets Market Fundamentals}}.{\BBCQ}
\newblock
\BIn{} M.~Prokopczuk\ (\BED), \APACrefbtitle {{Energy Pricing Models. Recent
  Advances, Methods, and Tools}} {{Energy Pricing Models. Recent Advances,
  Methods, and Tools}}\ (\BPGS\ 147--193).
\newblock
\begin{APACrefDOI} \doi{10.1007/978-1-137-37027-3\_6} \end{APACrefDOI}
\PrintBackRefs{\CurrentBib}

\bibitem [\protect \citeauthoryear {%
Hastie%
, Tibshirani%
\BCBL {}\ \BBA {} Friedman%
}{%
Hastie%
\ \protect \BOthers {.}}{%
{\protect \APACyear {2001}}%
}]{%
hastie2001}
\APACinsertmetastar {%
hastie2001}%
\begin{APACrefauthors}%
Hastie, T.%
, Tibshirani, R.%
\BCBL {}\ \BBA {} Friedman, J.%
\end{APACrefauthors}%
\unskip\
\newblock
\APACrefYear{2001}.
\newblock
\APACrefbtitle {The {Elements} of {Statistical} Learning} {The {Elements} of
  {Statistical} learning}.
\newblock
\APACaddressPublisher{}{Springer}.
\newblock
\begin{APACrefDOI} \doi{10.1007/978-0-387-84858-7} \end{APACrefDOI}
\PrintBackRefs{\CurrentBib}

\bibitem [\protect \citeauthoryear {%
Hochreiter%
\ \BBA {} Schmidhuber%
}{%
Hochreiter%
\ \BBA {} Schmidhuber%
}{%
{\protect \APACyear {1997}}%
}]{%
hochreiter1997}
\APACinsertmetastar {%
hochreiter1997}%
\begin{APACrefauthors}%
Hochreiter, S.%
\BCBT {}\ \BBA {} Schmidhuber, J.%
\end{APACrefauthors}%
\unskip\
\newblock
\APACrefYearMonthDay{1997}{}{}.
\newblock
{\BBOQ}\APACrefatitle {{Long Short-Term Memory}} {{Long Short-Term
  Memory}}.{\BBCQ}
\newblock
\APACjournalVolNumPages{Neural Computation}{9}{8}{1735--1780}.
\newblock
\begin{APACrefDOI} \doi{10.1162\%2Fneco.1997.9.8.1735} \end{APACrefDOI}
\PrintBackRefs{\CurrentBib}

\bibitem [\protect \citeauthoryear {%
Hornik%
}{%
Hornik%
}{%
{\protect \APACyear {1991}}%
}]{%
hornik1991}
\APACinsertmetastar {%
hornik1991}%
\begin{APACrefauthors}%
Hornik, K.%
\end{APACrefauthors}%
\unskip\
\newblock
\APACrefYearMonthDay{1991}{}{}.
\newblock
{\BBOQ}\APACrefatitle {Approximation capabilities of multilayer feedforward
  networks.} {Approximation capabilities of multilayer feedforward
  networks.}{\BBCQ}
\newblock
\APACjournalVolNumPages{Neural Networks}{4}{2}{251--257}.
\newblock
\begin{APACrefDOI} \doi{10.1016/0893-6080(91)90009-T} \end{APACrefDOI}
\PrintBackRefs{\CurrentBib}

\bibitem [\protect \citeauthoryear {%
Ioffe%
\ \BBA {} Szegedy%
}{%
Ioffe%
\ \BBA {} Szegedy%
}{%
{\protect \APACyear {2015}}%
}]{%
ioffe2015}
\APACinsertmetastar {%
ioffe2015}%
\begin{APACrefauthors}%
Ioffe, S.%
\BCBT {}\ \BBA {} Szegedy, C.%
\end{APACrefauthors}%
\unskip\
\newblock
\APACrefYearMonthDay{2015}{}{}.
\newblock
{\BBOQ}\APACrefatitle {Batch Normalization: Accelerating Deep Network Training
  by Reducing Internal Covariate Shift} {Batch normalization: Accelerating deep
  network training by reducing internal covariate shift}.{\BBCQ}
\newblock
\APACjournalVolNumPages{CoRR}{abs/1502.03167}{}{}.
\PrintBackRefs{\CurrentBib}

\bibitem [\protect \citeauthoryear {%
Keles%
, Scelle%
, Paraschiv%
\BCBL {}\ \BBA {} Fichtner%
}{%
Keles%
\ \protect \BOthers {.}}{%
{\protect \APACyear {2016}}%
}]{%
keles2016}
\APACinsertmetastar {%
keles2016}%
\begin{APACrefauthors}%
Keles, D.%
, Scelle, J.%
, Paraschiv, F.%
\BCBL {}\ \BBA {} Fichtner, W.%
\end{APACrefauthors}%
\unskip\
\newblock
\APACrefYearMonthDay{2016}{01}{}.
\newblock
{\BBOQ}\APACrefatitle {{Extended forecast methods for day-ahead electricity
  prices applying artificial neural networks}} {{Extended forecast methods for
  day-ahead electricity prices applying artificial neural networks}}.{\BBCQ}
\newblock
\APACjournalVolNumPages{Applied Energy}{162}{}{218--230}.
\newblock
\begin{APACrefDOI} \doi{10.1016/j.apenergy.2015.09.087} \end{APACrefDOI}
\PrintBackRefs{\CurrentBib}

\bibitem [\protect \citeauthoryear {%
Kingma%
\ \BBA {} Ba%
}{%
Kingma%
\ \BBA {} Ba%
}{%
{\protect \APACyear {2015}}%
}]{%
kingma2015}
\APACinsertmetastar {%
kingma2015}%
\begin{APACrefauthors}%
Kingma, D\BPBI P.%
\BCBT {}\ \BBA {} Ba, J.%
\end{APACrefauthors}%
\unskip\
\newblock
\APACrefYearMonthDay{2015}{}{}.
\newblock
{\BBOQ}\APACrefatitle {Adam: {A} Method for Stochastic Optimization} {Adam: {A}
  method for stochastic optimization}.{\BBCQ}
\newblock
\BIn{} Y.~Bengio\ \BBA {} Y.~LeCun\ (\BEDS), \APACrefbtitle {3rd International
  Conference on Learning Representations, {ICLR} 2015, San Diego, CA, USA, May
  7-9, 2015, Conference Track Proceedings.} {3rd international conference on
  learning representations, {ICLR} 2015, san diego, ca, usa, may 7-9, 2015,
  conference track proceedings.}
\newblock
\begin{APACrefURL} \url{http://arxiv.org/abs/1412.6980} \end{APACrefURL}
\PrintBackRefs{\CurrentBib}

\bibitem [\protect \citeauthoryear {%
Lago%
, Ridder%
\BCBL {}\ \BBA {} Schutter%
}{%
Lago%
\ \protect \BOthers {.}}{%
{\protect \APACyear {2018}}%
}]{%
lago2018}
\APACinsertmetastar {%
lago2018}%
\begin{APACrefauthors}%
Lago, J.%
, Ridder, F\BPBI D.%
\BCBL {}\ \BBA {} Schutter, B\BPBI D.%
\end{APACrefauthors}%
\unskip\
\newblock
\APACrefYearMonthDay{2018}{}{}.
\newblock
{\BBOQ}\APACrefatitle {{Forecasting spot electricity prices: Deep learning
  approaches and empirical comparison of traditional algorithms}} {{Forecasting
  spot electricity prices: Deep learning approaches and empirical comparison of
  traditional algorithms}}.{\BBCQ}
\newblock
\APACjournalVolNumPages{Applied Energy}{221}{}{386--405}.
\newblock
\begin{APACrefDOI} \doi{10.1016/j.apenergy.2018.02.069} \end{APACrefDOI}
\PrintBackRefs{\CurrentBib}

\bibitem [\protect \citeauthoryear {%
Marcjasz%
, Uniejewski%
\BCBL {}\ \BBA {} Weron%
}{%
Marcjasz%
\ \protect \BOthers {.}}{%
{\protect \APACyear {2018}}%
}]{%
marcjasz2018}
\APACinsertmetastar {%
marcjasz2018}%
\begin{APACrefauthors}%
Marcjasz, G.%
, Uniejewski, B.%
\BCBL {}\ \BBA {} Weron, R.%
\end{APACrefauthors}%
\unskip\
\newblock
\APACrefYearMonthDay{2018}{}{}.
\newblock
{\BBOQ}\APACrefatitle {{On the importance of the long-term seasonal component
  in day-ahead electricity price forecasting with NARX neural networks}} {{On
  the importance of the long-term seasonal component in day-ahead electricity
  price forecasting with NARX neural networks}}.{\BBCQ}
\newblock
\APACjournalVolNumPages{International Journal of Forecasting}{}{}{}.
\newblock
\begin{APACrefDOI} \doi{10.1016/j.ijforecast.2017.11.009} \end{APACrefDOI}
\PrintBackRefs{\CurrentBib}

\bibitem [\protect \citeauthoryear {%
Mosbah%
\ \BBA {} El-Hawary%
}{%
Mosbah%
\ \BBA {} El-Hawary%
}{%
{\protect \APACyear {2016}}%
}]{%
mosbah2016}
\APACinsertmetastar {%
mosbah2016}%
\begin{APACrefauthors}%
Mosbah, H.%
\BCBT {}\ \BBA {} El-Hawary, M\BPBI E.%
\end{APACrefauthors}%
\unskip\
\newblock
\APACrefYearMonthDay{2016}{}{}.
\newblock
{\BBOQ}\APACrefatitle {{Hourly Electricity Price Forecasting for the Next Month
  Using Multilayer Neural Network}} {{Hourly Electricity Price Forecasting for
  the Next Month Using Multilayer Neural Network}}.{\BBCQ}
\newblock
\APACjournalVolNumPages{Canadian Journal of Electrical and Computer
  Engineering}{39}{}{283-291}.
\PrintBackRefs{\CurrentBib}

\bibitem [\protect \citeauthoryear {%
Nogales%
, Contreras%
, J.~Conejo%
\BCBL {}\ \BBA {} Espinola%
}{%
Nogales%
\ \protect \BOthers {.}}{%
{\protect \APACyear {2002}}%
}]{%
nogales2002}
\APACinsertmetastar {%
nogales2002}%
\begin{APACrefauthors}%
Nogales, F.%
, Contreras, J.%
, J.~Conejo, A.%
\BCBL {}\ \BBA {} Espinola, R.%
\end{APACrefauthors}%
\unskip\
\newblock
\APACrefYearMonthDay{2002}{04}{}.
\newblock
{\BBOQ}\APACrefatitle {{Forecasting Next-Day Electricity Prices by Time Series
  Models}} {{Forecasting Next-Day Electricity Prices by Time Series
  Models}}.{\BBCQ}
\newblock
\APACjournalVolNumPages{Power Engineering Review, IEEE}{22}{}{58--58}.
\newblock
\begin{APACrefDOI} \doi{10.1109/MPER.2002.4312063} \end{APACrefDOI}
\PrintBackRefs{\CurrentBib}

\bibitem [\protect \citeauthoryear {%
Srivastava%
, Hinton%
, Krizhevsky%
, Sutskever%
\BCBL {}\ \BBA {} Salakhutdinov%
}{%
Srivastava%
\ \protect \BOthers {.}}{%
{\protect \APACyear {2014}}%
}]{%
srivastava2014}
\APACinsertmetastar {%
srivastava2014}%
\begin{APACrefauthors}%
Srivastava, N.%
, Hinton, G.%
, Krizhevsky, A.%
, Sutskever, I.%
\BCBL {}\ \BBA {} Salakhutdinov, R.%
\end{APACrefauthors}%
\unskip\
\newblock
\APACrefYearMonthDay{2014}{{\APACmonth{01}}}{}.
\newblock
{\BBOQ}\APACrefatitle {Dropout: A Simple Way to Prevent Neural Networks from
  Overfitting} {Dropout: A simple way to prevent neural networks from
  overfitting}.{\BBCQ}
\newblock
\APACjournalVolNumPages{J. Mach. Learn. Res.}{15}{1}{1929--1958}.
\newblock
\begin{APACrefURL} \url{http://dl.acm.org/citation.cfm?id=2627435.2670313}
  \end{APACrefURL}
\PrintBackRefs{\CurrentBib}

\bibitem [\protect \citeauthoryear {%
Thieleman%
\ \BBA {} Hinton%
}{%
Thieleman%
\ \BBA {} Hinton%
}{%
{\protect \APACyear {2012}}%
}]{%
tieleman2012}
\APACinsertmetastar {%
tieleman2012}%
\begin{APACrefauthors}%
Thieleman, T.%
\BCBT {}\ \BBA {} Hinton, G.%
\end{APACrefauthors}%
\unskip\
\newblock
\APACrefYearMonthDay{2012}{}{}.
\newblock
\APACrefbtitle {Lecture 6.5 -- rmsprop: Divide the gradient by a running
  average of its recent magnitude.} {Lecture 6.5 -- rmsprop: Divide the
  gradient by a running average of its recent magnitude.}
\PrintBackRefs{\CurrentBib}

\bibitem [\protect \citeauthoryear {%
Troncoso%
, Santos%
, Gomez-Exposito%
, Martinez-Ramos%
\BCBL {}\ \BBA {} Riquelme%
}{%
Troncoso%
\ \protect \BOthers {.}}{%
{\protect \APACyear {2007}}%
}]{%
troncoso2007}
\APACinsertmetastar {%
troncoso2007}%
\begin{APACrefauthors}%
Troncoso, A.%
, Santos, J.%
, Gomez-Exposito, A.%
, Martinez-Ramos, J.%
\BCBL {}\ \BBA {} Riquelme, J.%
\end{APACrefauthors}%
\unskip\
\newblock
\APACrefYearMonthDay{2007}{09}{}.
\newblock
{\BBOQ}\APACrefatitle {{Electricity Market Price Forecasting Based on Weighted
  Nearest Neighbors Techniques}} {{Electricity Market Price Forecasting Based
  on Weighted Nearest Neighbors Techniques}}.{\BBCQ}
\newblock
\APACjournalVolNumPages{IEEE Transactions on Power Systems}{22}{}{1294--1301}.
\newblock
\begin{APACrefDOI} \doi{10.1109/TPWRS.2007.901670} \end{APACrefDOI}
\PrintBackRefs{\CurrentBib}

\bibitem [\protect \citeauthoryear {%
Wagner%
}{%
Wagner%
}{%
{\protect \APACyear {2014}}%
}]{%
wagner2014}
\APACinsertmetastar {%
wagner2014}%
\begin{APACrefauthors}%
Wagner, A.%
\end{APACrefauthors}%
\unskip\
\newblock
\APACrefYearMonthDay{2014}{}{}.
\newblock
{\BBOQ}\APACrefatitle {Residual Demand Modeling and Application to Electricity
  Pricing} {Residual demand modeling and application to electricity
  pricing}.{\BBCQ}
\newblock
\APACjournalVolNumPages{The Energy Journal}{35}{2}{45--73}.
\newblock
\begin{APACrefDOI} \doi{10.2139/ssrn.2018908} \end{APACrefDOI}
\PrintBackRefs{\CurrentBib}

\bibitem [\protect \citeauthoryear {%
Weron%
}{%
Weron%
}{%
{\protect \APACyear {2014}}%
}]{%
weron2014}
\APACinsertmetastar {%
weron2014}%
\begin{APACrefauthors}%
Weron, R.%
\end{APACrefauthors}%
\unskip\
\newblock
\APACrefYearMonthDay{2014}{}{}.
\newblock
{\BBOQ}\APACrefatitle {{Electricity price forecasting: A review of the
  state-of-the-art with a look into the future}} {{Electricity price
  forecasting: A review of the state-of-the-art with a look into the
  future}}.{\BBCQ}
\newblock
\APACjournalVolNumPages{International Journal of
  Forecasting}{30}{4}{1030--1081}.
\newblock
\begin{APACrefDOI} \doi{10.1016/j.ijforecast.2014.08.008} \end{APACrefDOI}
\PrintBackRefs{\CurrentBib}

\bibitem [\protect \citeauthoryear {%
Würzburg%
, Labandeira%
\BCBL {}\ \BBA {} Linares%
}{%
Würzburg%
\ \protect \BOthers {.}}{%
{\protect \APACyear {2013}}%
}]{%
wuerzburg2013}
\APACinsertmetastar {%
wuerzburg2013}%
\begin{APACrefauthors}%
Würzburg, K.%
, Labandeira, X.%
\BCBL {}\ \BBA {} Linares, P.%
\end{APACrefauthors}%
\unskip\
\newblock
\APACrefYearMonthDay{2013}{}{}.
\newblock
{\BBOQ}\APACrefatitle {{Renewable generation and electricity prices: Taking
  stock and new evidence for Germany and Austria}} {{Renewable generation and
  electricity prices: Taking stock and new evidence for Germany and
  Austria}}.{\BBCQ}
\newblock
\APACjournalVolNumPages{Energy Economics}{40}{S1}{159--171}.
\newblock
\begin{APACrefDOI} \doi{10.1016/j.eneco.2013.09.011} \end{APACrefDOI}
\PrintBackRefs{\CurrentBib}

\bibitem [\protect \citeauthoryear {%
Ziel%
\ \BBA {} Steinert%
}{%
Ziel%
\ \BBA {} Steinert%
}{%
{\protect \APACyear {2016}}%
}]{%
ziel2016}
\APACinsertmetastar {%
ziel2016}%
\begin{APACrefauthors}%
Ziel, F.%
\BCBT {}\ \BBA {} Steinert, R.%
\end{APACrefauthors}%
\unskip\
\newblock
\APACrefYearMonthDay{2016}{}{}.
\newblock
{\BBOQ}\APACrefatitle {{Electricity Price Forecasting using Sale and Purchase
  Curves: The X-Model}} {{Electricity Price Forecasting using Sale and Purchase
  Curves: The X-Model}}.{\BBCQ}
\newblock
\APACjournalVolNumPages{Energy Economics}{59}{}{435--454}.
\newblock
\begin{APACrefDOI} \doi{10.1016/j.eneco.2016.08.008} \end{APACrefDOI}
\PrintBackRefs{\CurrentBib}

\bibitem [\protect \citeauthoryear {%
Ziel%
, Steinert%
\BCBL {}\ \BBA {} Husmann%
}{%
Ziel%
\ \protect \BOthers {.}}{%
{\protect \APACyear {2015}}%
}]{%
ziel2015}
\APACinsertmetastar {%
ziel2015}%
\begin{APACrefauthors}%
Ziel, F.%
, Steinert, R.%
\BCBL {}\ \BBA {} Husmann, S.%
\end{APACrefauthors}%
\unskip\
\newblock
\APACrefYearMonthDay{2015}{}{}.
\newblock
{\BBOQ}\APACrefatitle {{Forecasting day ahead electricity spot prices: The
  impact of the EXAA to other European electricity markets}} {{Forecasting day
  ahead electricity spot prices: The impact of the EXAA to other European
  electricity markets}}.{\BBCQ}
\newblock
\APACjournalVolNumPages{Energy Economics}{51}{}{430--444}.
\newblock
\begin{APACrefDOI} \doi{10.1016/j.eneco.2015.08.005} \end{APACrefDOI}
\PrintBackRefs{\CurrentBib}

\bibitem [\protect \citeauthoryear {%
Ziel%
\ \BBA {} Weron%
}{%
Ziel%
\ \BBA {} Weron%
}{%
{\protect \APACyear {2018}}%
}]{%
ziel2018}
\APACinsertmetastar {%
ziel2018}%
\begin{APACrefauthors}%
Ziel, F.%
\BCBT {}\ \BBA {} Weron, R.%
\end{APACrefauthors}%
\unskip\
\newblock
\APACrefYearMonthDay{2018}{}{}.
\newblock
{\BBOQ}\APACrefatitle {{Day-ahead electricity price forecasting with
  high-dimensional structures: Univariate vs. multivariate modeling
  frameworks}} {{Day-ahead electricity price forecasting with high-dimensional
  structures: Univariate vs. multivariate modeling frameworks}}.{\BBCQ}
\newblock
\APACjournalVolNumPages{Energy Economics}{70}{}{396--420}.
\newblock
\begin{APACrefDOI} \doi{10.1016/j.eneco.2017.12.016} \end{APACrefDOI}
\PrintBackRefs{\CurrentBib}

\end{thebibliography}
\end{document}